\documentclass{llncs}
\usepackage{graphicx}
\usepackage[caption=false]{subfig}
\usepackage{hyperref}
\usepackage{booktabs}
\usepackage{amsfonts}
\usepackage{booktabs}
\usepackage{siunitx}
\usepackage{multirow} 
\usepackage{placeins}
\usepackage{afterpage}
\usepackage[utf8x]{inputenc}
\begin{document}
\title{Heterogeneity and Instability in the Stable Marriage Problem}
\titlerunning{HISMP}  
\author{Bernardo Alves Furtado\inst{1,2}}
\authorrunning{Furtado, B.} 
\tocauthor{Bernardo Alves Furtado}
\institute{Institute for Applied Economic Research (IPEA), Brazil\\
\email{bernardo.furtado at ipea.gov.br},\\ 
\and
National Council of Research (CNPq)}
\maketitle              
\begin{abstract}
The Stable Marriage Problem (SMP) has been extremely discussed in the literature and it is useful to a number of real-world applications. We propose a generalized version of the SMP in which numbers of the matching groups are different as in \cite{shi_instability_2018}. However, we go further to make a percentage of each group behave as active message senders. As such, the special case in which all Males are active messengers ($beta = 1$) and all Females are not active ($alpha = 0$) replicates the results in \cite{shi_instability_2018}. Moreover, we use numerical simulation to present three cases (and their extremes) in which we vary the percentage of active messengers in each group. Whereas we are able to replicate previous work, our numerical simulations also suggest that socially optimal comes only when the groups are homogeneous. More real-world like results are presented when members from both groups are active message senders. 

\keywords{Stable Marriage Problem, Heterogeneity, Model Generalization}
\end{abstract}

\section{Introduction}
The Stable Marriage Problem (SMP) proposed by Gale and Shapley \cite{gale_college_1962} matches members from two distinct groups - usually Males and Females - in which, each member of each group has all the members of the other group ranked ordinarily in degree of preference. Members of one group (Males) are active when sending messages (with proposals) to members of the other group (Females). Although not having actually initiated the contact, the receiving end (Females) always compares with current partner (if any) ranking and always chooses the highest ranked bid, discarding previous match if necessary. The problem has been shown to have a stable solution \cite{shi_instability_2018,mcvitie_stable_1971,dotsenko_exact_2000}.

Since the SMP proposal, the algorithm has been debated, extended and discussed in the literature \cite{mcvitie_stable_1971,roth_evolution_1984}. Recently, SMP has been applied to new economics transport apps \cite{kummel_taxi_2016}, to device communication and the internet \cite{xiao_enhance_2016,hasan_distributed_2016} and actual partner matching problems within internet apps \cite{hitsch_matching_2010}.

Most of the literature, however, focuses on SMP with both groups of members being of equal size. Shi et al. \cite{shi_instability_2018} just recently proposed a generalization in which the number of the receiving agents (Females) are varied - from nearly none to double the size of the number of members on the other group (Males), advancing on previous unequal group-sized work \cite{dzierzawa_statistics_2000}. Shi et al. shows that the stable solution is very sensitive to the number of agents and happiness of members from each group change dramatically when departing from their equal numbers. Thus, suggesting the instability of the SMP.

We propose to further generalize the SMP problem in which we have varying sizes of one of the groups (Females) and thus the Instability in the SMP, namely ISMP. We go further to propose that not only members of one group may be active and send messages, but a varying percentage of each group and their combinations can act as the active end as well.

Our results replicate previous work \cite{shi_instability_2018} and suggest that the stability is very attached to the full membership of all members to a choice of being active or passive (in its entirety). Small deviations, such as a .1 percent of members from the Female group becoming active message senders changes the equilibrium. In most cases, these new results are less socially optimal than the homogeneous ones. 

Besides this short introduction, we present the method, the results of the numerical simulation of the three cases. We conclude with discussions.

\section{Methods}

Initially, we follow the traditional procedures summarized in \cite{shi_instability_2018}.

\begin{enumerate}
    \item $N$ Male and $M$ Female seek to match with a partner.
    \item Each individual has a personal, random ranking of preference, ordering all members of the opposite group.
    \item Energy, or satisfaction, of each member is equivalent to the position of their partner in their own ranking. Note, the highest the ranking, the lower is the energy. Thus, the lowest energy is preferable. 
    \item Singles' energy is given by the length of the opposite group, plus one, following \cite{shi_instability_2018}.
    \item The strong condition to solve the algorithm is that each run in the simulation only finishes when all active messengers have sent messages to all possible recipients. Thus, it is guaranteed that every single active messenger has been rejected by all available recipients of the other group.
    \item The recipient members always change partner for the better when receiving proposals.
\end{enumerate}
As it is shown in the description above, we have changed members of a given group (Males or Females) into active (or not members). In order to implement that, the procedures are as follows:

\begin{enumerate}
    \item We introduce two parameters which are the probabilistic percentage of active members in that group. Hence, the replication of the previous procedure would have $\alpha = 0$ for no active Female members and $\beta = 1$ for all Male members to be active message senders.
    \item Further, instead of having all Male members been rejected before the iteration, we apply the strong condition that \texttt{all active members who remain single have been rejected by all members of the opposite group}.
        
\end{enumerate}

Those procedures guarantee that when we run the numerical simulation we are varying (a) the size of one of the groups (Females), whereas maintaining the other fixed in size (1,000 Males), but also (b) the percentage of active messengers in each group. For the results presented here, we repeated each simulation 50 times.

Systematically, we study three cases: 
\begin{itemize}
    \item \textbf{Case 1.} In which all Males are always active ($\beta = 1$) and we vary the number of active Females ($\alpha \in [0, 1]$)
    \item \textbf{Case 2.} In which active messengers probabilities percentages add to 1, so that Males ($\beta \in [0, 1]$) and Females ($\alpha = 1 - \beta$)
    \item \textbf{Case 3.} In which all Females are always active ($\alpha = 1$) and we vary the number of active Males ($\beta \in [0, 1]$)
\end{itemize}

The code repository (in python) is available at \href{https://github.com/BAFurtado/HISMP}{GitHub.com/BAFurtado/HISMP},

\section{Results and Discussion}

In this section we discuss the results of the three cases which aim at covering the range of possible probability percentage of active members from both groups: Males and Females. 

\subsection{Special Cases: extremes}

\begin{figure}[ht]
\centering
\subfloat[$\alpha = 1, \beta = 1$]{{\includegraphics[width=5.7cm]{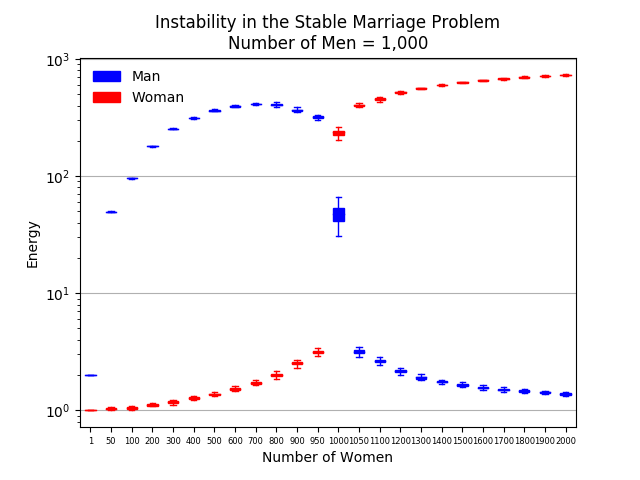}}}
\qquad
\subfloat[$\alpha = 1, \beta = 0$]{{\includegraphics[width=5.7cm]{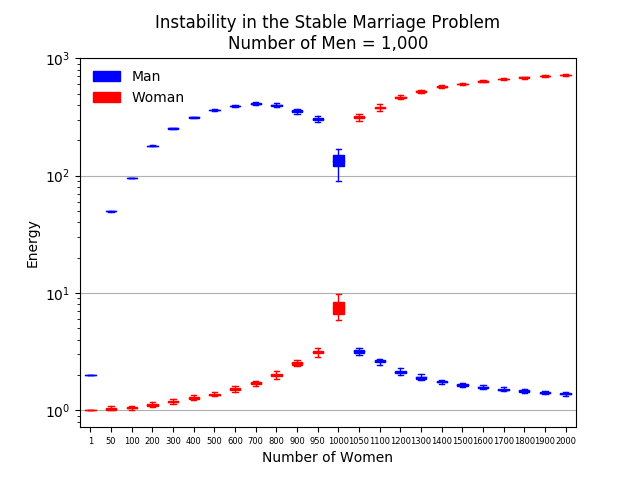}}}
\qquad
\subfloat[$\alpha = 0, \beta = 1$]{{\includegraphics[width=5.7cm]{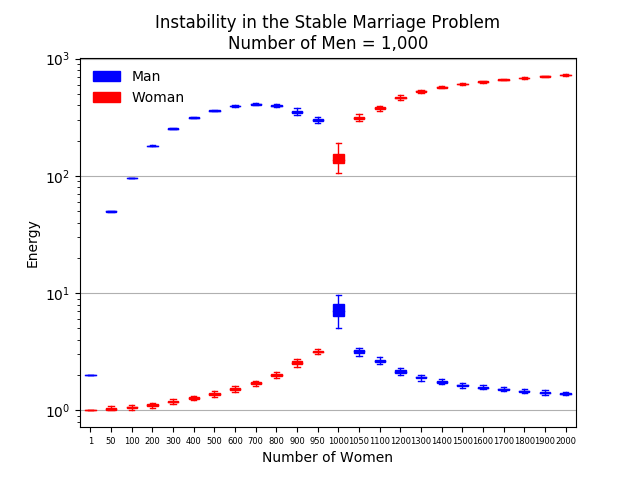}}}
\qquad
\subfloat[$\alpha = .5, \beta = .5$]{{\includegraphics[width=5.7cm]{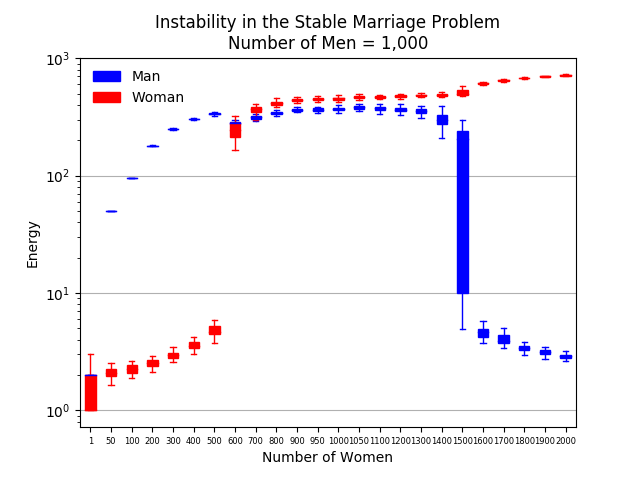}}}
\caption{This figure shows the extreme cases in which $\alpha$ and $\beta$ are 1 or 0 and when they are both .5.}
\label{fig:Special cases}
\end{figure}

First and foremost, we discuss the extreme cases. Those in which either the parameters is zero, or the case when their probability is the same (Fig. \ref{fig:Special cases}). 

The interesting result is that the energy of each of the groups through the variation in size of Females is very similar when all Males are active (and no Females), in the opposite case, and when both Males and Females are active. Note that Fig. \ref{fig:Special cases} (c) is the replication of the results obtained in \cite{shi_instability_2018}. 

The slight change happens when Males are active and the groups are of the same size, Males have lower energy. Being the opposite when Females are active. 

When half of the population of each group is active messengers (Fig. \ref{fig:Special cases} (c)), the results suggest that on average both groups are worth off when comparing to the other three cases shown. This seems to be relevant as it indicates that more active individual-seeking behavior results in lower than socially optimal benefits. 

\subsection{\textbf{Case 1}: males are always active: $\beta = 1$, $\alpha = [0, 1]$}

\begin{figure}[ht]
\centering
\subfloat[$\alpha = .1$, $\beta = 1$]{{\includegraphics[width=5.7cm]{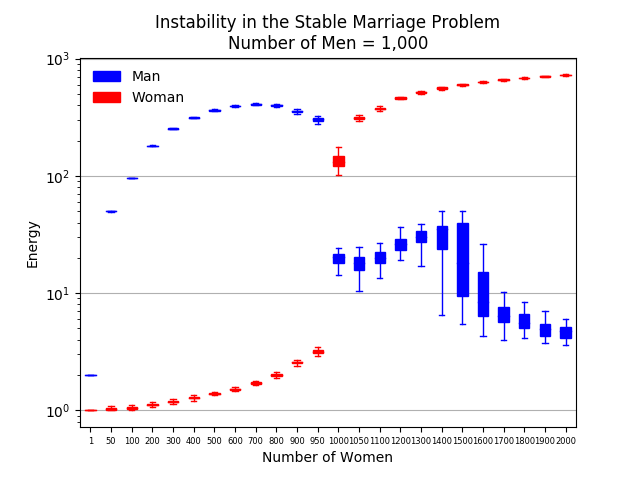}}}
\qquad
\subfloat[$\alpha = .4$, $\beta = 1$]{{\includegraphics[width=5.7cm]{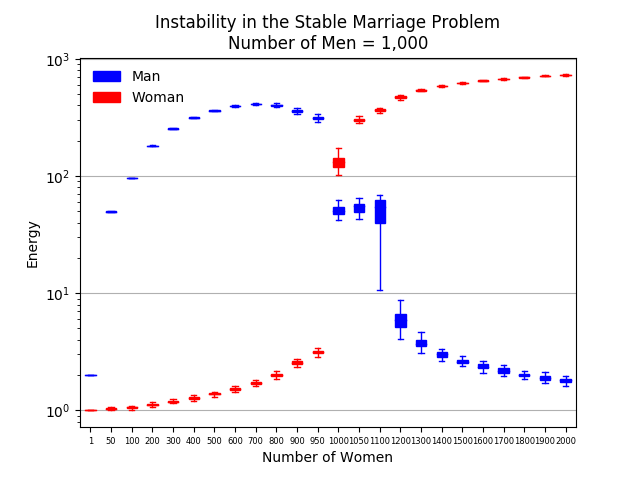}}}
\qquad
\subfloat[$\alpha = .6$, $\beta = 1$]{{\includegraphics[width=5.7cm]{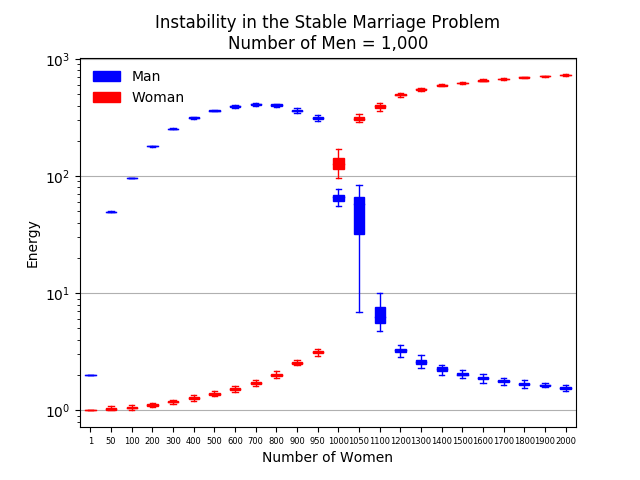}}}
\qquad
\subfloat[$\alpha = .9$, $\beta = 1$]{{\includegraphics[width=5.7cm]{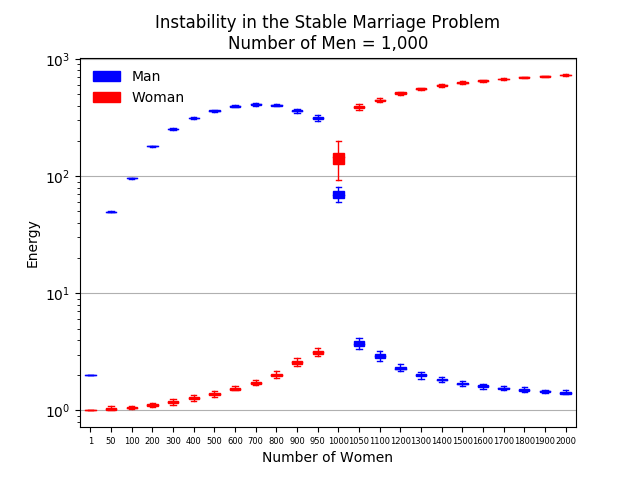}}}
\caption{This figure shows \textbf{Case 1} when males are always active: $\beta = 1$, $\alpha = [0, 1]$}
\label{fig:Case1}
\end{figure}

The results of \textbf{Case 1} (Fig. \ref{fig:Case1}) are also of interest because it clearly shows that active messaging from the Females group actually interferes only with the energy (satisfaction) of those on the other group. Further, the slightest of the percentage of Females that are active (.1) results in the highest disturbance and loss of energy in the group of Males. That suggests that a minimum parcel of the opposite group becomes active in the matching game, large interference it imposes in the general equilibrium.

\subsection{\textbf{Case 2}: active agents vary: $\beta = [0, 1]$, $\alpha = 1 - \beta$}

\begin{figure}[ht]
\centering
\subfloat[$\alpha = .1$,  $\beta = .9$]{{\includegraphics[width=5.7cm]{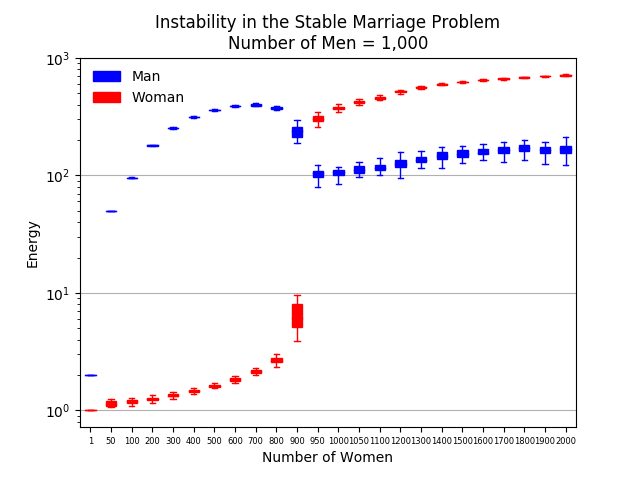}}}
\qquad
\subfloat[$\alpha = .4$,  $\beta = .6$]{{\includegraphics[width=5.7cm]{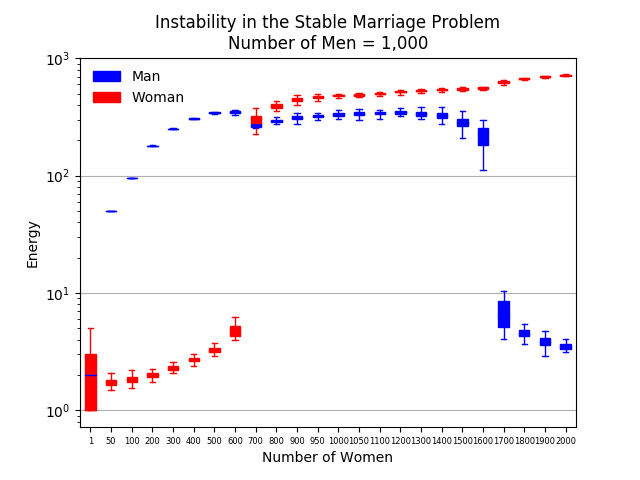}}}
\qquad
\subfloat[$\alpha = .6$,  $\beta = .4$]{{\includegraphics[width=5.7cm]{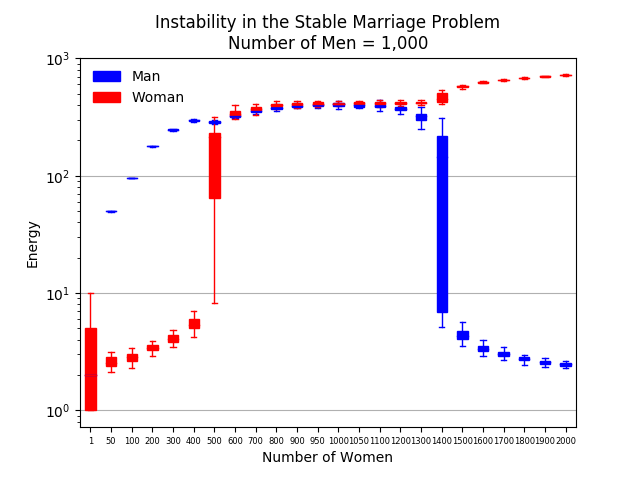}}}
\qquad
\subfloat[$\alpha = .9$,  $\beta = .1$]{{\includegraphics[width=5.7cm]{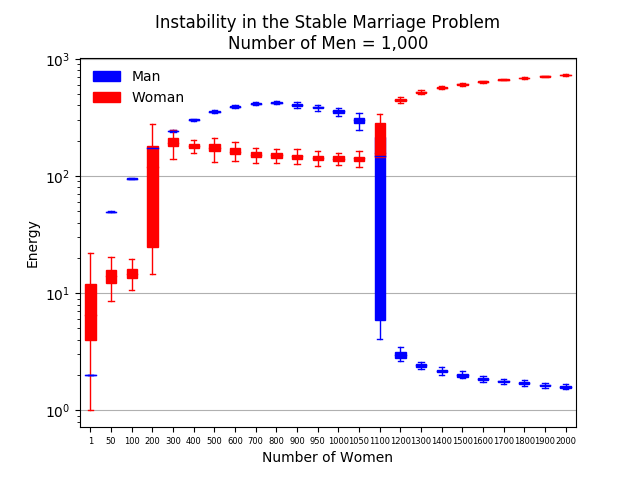}}}
\caption{This figure shows \textbf{Case 2} when active agents vary: $\beta = [0, 1]$, $\alpha = 1 - \beta$}
\label{fig:Case2}
\end{figure}

This case probably represents the most likely real-case scenario. What Fig. \ref{fig:Case2} depicts is that stability is far from socially optimal (as observed in Fig. \ref{fig:Special cases}) and that the group with the highest percentage is less successful in bringing their energy to better level when compared to the other group. Thus, Females show their best results when they have the least percentage (.1), compared to Males at .9. The opposite happens when Feales have their highest percentage of active members (.9) and Males have their smallest (.1).

\subsection{\textbf{Case 3}: females are always active: $\beta = [0, 1]$, $\alpha = 1$}

\begin{figure}[ht]
\centering
\subfloat[$\alpha = 1$,  $\beta = .1$]{{\includegraphics[width=5.7cm]{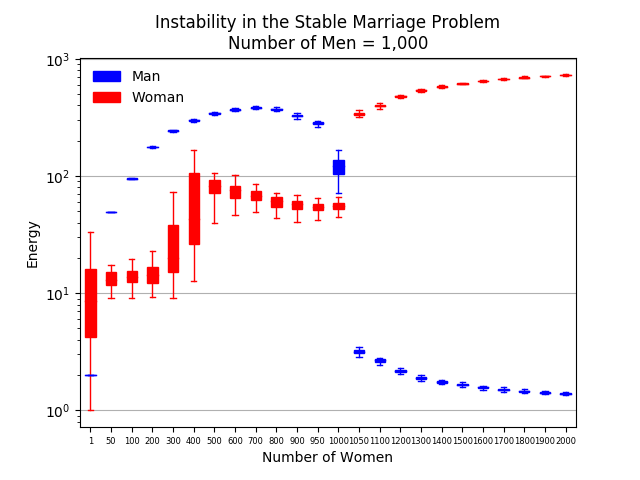}}}
\qquad
\subfloat[$\alpha = 1$,  $\beta = .4$]{{\includegraphics[width=5.7cm]{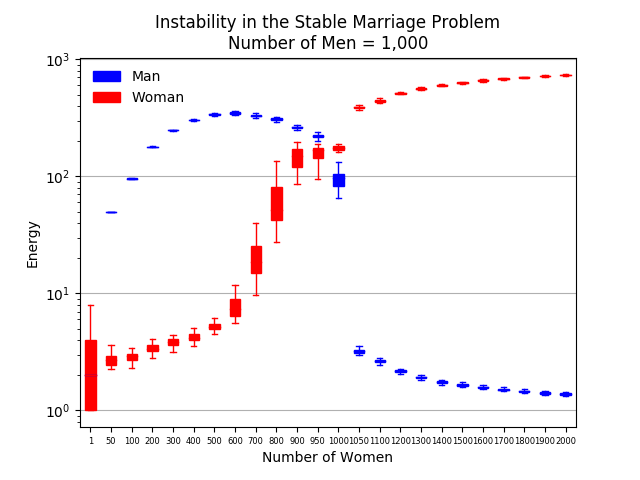}}}
\qquad
\subfloat[$\alpha = 1$, $\beta = .6$]{{\includegraphics[width=5.7cm]{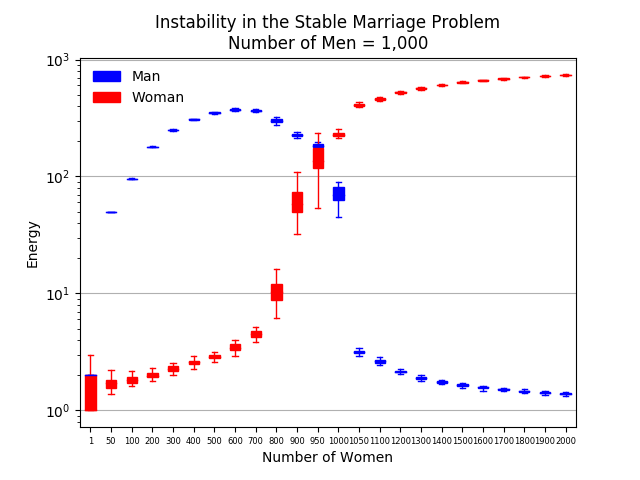}}}
\qquad
\subfloat[$\alpha = 1$, $\beta = .9$]{{\includegraphics[width=5.7cm]{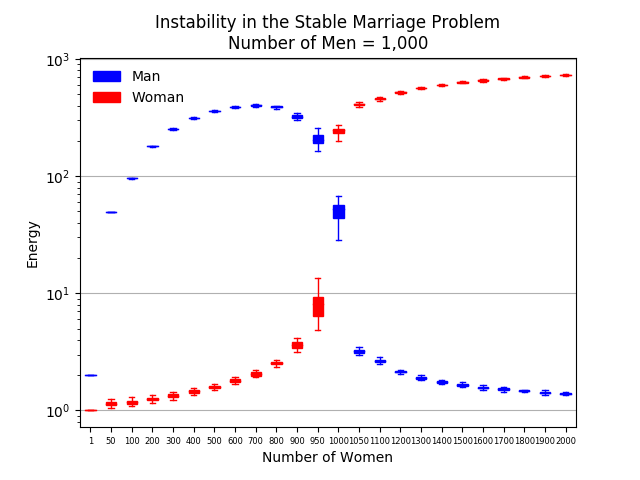}}}
\caption{This figure shows \textbf{Case 3} females are always active: $\beta = [0, 1]$, $\alpha = 1$}
\label{fig:Case3}
\end{figure}

In \textbf{Case 3} (Fig. \ref{fig:Case3}), similarly to what happens in \textbf{Case 1}, only Female energy is affected when we change the probability of Male active messengers and keep all Females as active. The best results for Females happens when Males approach the full probability of becoming active. On the other hand, Females energy is more deeply disturbed when just a fraction (.1) of Males are active.

\section{Conclusion}

In this paper we further generalize the SMP problem varying the size of the matching groups, as in \cite{shi_instability_2018}, but also further changing the probability of active messengers in each group. 

Results suggest that early, optimal results of the SMP remain strongly valid for the three cases in which all members of both groups are active or members of just one group is active. Real applications, however, are much more likely to encounter situations in which there are members in both groups that actively seed to propose (engage) with members of the other group. In such cases, as observed in \textbf{Case 2}, results are also stable, but definitely less socially optimal when compared to \textbf{just one group active} cases. 

Although, theoretically, when can correctly argue that active members are in essence the same, their percentage presence in each group may vary. Hence, the fact of belonging to a given group and having the members of the other group as targets for matching is relevant. 

Future work may propose to further generalize the problem in which a portion of each group may try to seek matching within their own group. 

\section{Acknowledgments}
The author would like to thank early comments from Adrián Carro on the generalization of agents (males and females) and pontual early help with the code generalization from Jéssica Fernandes de Araújo. Further, we acknowledge Grant
[306954/2016-8] from the National Council of Research
(CNPq) and a sabbatical leave from the Institute for Applied Economic Research (Ipea).


\end{document}